\documentclass[conference]{IEEEtran}
\IEEEoverridecommandlockouts
\usepackage{hyperref}
\usepackage{cite}
\usepackage{amsmath,amssymb,amsfonts}
\usepackage{algpseudocode}
\usepackage{algorithm}
\usepackage{authblk}
\usepackage{graphicx}
\usepackage{textcomp}
\usepackage{xcolor}
\usepackage{subcaption}
\usepackage[top=0.7in, left=0.673in, right=0.7in, bottom=1.013in]{geometry}
\def\BibTeX{{\rm B\kern-.05em{\sc i\kern-.025em b}\kern-.08em
    T\kern-.1667em\lower.7ex\hbox{E}\kern-.125emX}}
\begin{document}

\title{Deep Multi-modal Neural Receiver for 6G Vehicular Communication}

\author[ ]{
\begin{minipage}[t]{\textwidth}
\centering
Osama Saleem$^{1,2}$, Mohammed Alfaqawi$^{2}$, Pierre Merdrignac$^{2}$, 
Abdelaziz Bensrhair$^{1}$, Soheyb Ribouh$^{1}$
\end{minipage}
}
\affil[ ] {
\begin{minipage}[t]{\textwidth}
\centering
\textsuperscript{1}INSA Rouen Normandie, Univ Rouen Normandie, Université Le Havre Normandie, Normandie Univ, LITIS UR 4108, F-76000 Rouen, France.
\end{minipage}
}
\affil[2]{Institut VEDECOM, 23 bis Allée des Marronniers, 78000 Versailles, France}

\affil[ ]{
\begin{minipage}[t]{\textwidth}
\centering
Email: osama.saleem@insa-rouen.fr, 
\end{minipage}}





\maketitle

\begin{abstract}
Deep Learning (DL) based neural receiver models are used to jointly optimize PHY of baseline receiver for cellular vehicle to everything (C-V2X) system in next generation (6G) communication, however,  there has been no exploration of how varying training parameters affect the model's efficiency. Additionally, a comprehensive evaluation of its performance on multi-modal data remains largely unexplored. To address this, we propose a neural receiver designed to optimize Bit Error Rate (BER) for vehicle to network (V2N) uplink scenario in 6G network. We train multiple neural receivers by changing its trainable parameters and use the best fit model as proposition for large scale deployment. Our proposed neural receiver gets signal in frequency domain at the base station (BS) as input and generates optimal log likelihood ratio (LLR) at the output. It estimates the channel based on the received signal, equalizes and demodulates the higher order modulated signal. Later, to evaluate multi-modality of the proposed model, we test it across diverse V2X data flows (e.g., image, video, gps, lidar cloud points and radar detection signal). Results from simulation clearly indicates that our proposed multi-modal neural receiver outperforms state-of-the-art receiver architectures by achieving high performance at low Signal to Noise Ratio (SNR).
\end{abstract}


\section{Introduction}

The International Mobile Telecommunications (IMT) framework for 2030 has introduced advanced use cases for 6G, enabling technologies like massive and immersive communication, hyper reliable and low latency communication \cite{1}. In this context, Vehicle-to-Everything (V2X) communication is a key application in the automotive industry, addressing various safety and performance-critical use cases \cite{3} \cite{53}. To support automotive requirements, V2X communication deals with data from various sources such as lidar, radar, image, gps and audio \cite{52}. As an example, lidar, radar and image contribute to environmental awareness, GPS provide vehicle positioning and audio is used for human-machine interaction. Considering this multi-modal nature of the V2X data streams, one major challenge lies is ensuring high reliability for reconstructing signal from any data source. Here, we especially focus in minimizing the error at the base station for Vehicle to Network (V2N) uplink scenario, as any error in signal reception could compromise performance of the system \cite{22} \cite{2}.
Additionally, the 5G Automotive Association (5GAA) have also emphasized the importance of reliability in V2X communication, particularly in safety-critical use cases \cite{4}. 

Traditional communication system may fail to provide end to end performance in such scenarios. Such approach typically optimizes different receiver functionalities independently, which leads to suboptimal performance and increased error rates \cite{5}. In contrast, an AI based solution has the capability to provide efficient end to end performance in complex communication scenarios \cite{55}. By leveraging deep learning algorithms, it can efficiently manage the interference and complexity associated with diverse data types, reducing the Bit Error Rate (BER) during reconstruction of information bits. This has also been motivated by recent declaration of 3GPP to integrate AI into 6G cellular communication \cite{10}. In recent years, research on neural receivers have shown how machine learning can significantly enhance signal processing capabilities in wireless communication system.
In \cite{6}, the authors propose DeepRx, a convolution based architecture that jointly optimizes the receiver functionalities and is designed for 5G data symbols. DeepRx optimizes BER by replacing baseline PHY layer of receiver. Results from simulation show that DeepRx outperforms the conventional independant optimization of each functionality in term of error rate. Furthermore, \cite{7} propose TransRx, an attention based architecture that outperforms DeepRx in terms of reliability (Error rate). TransRx is designed specifically for 6G V2N communication scenario.
However, both of these architectures donot take into account how the variable training parameters (e.g., number of blocks, number of attention heads/convolution channels) can impact the performance. Additionally, evaluation of neural receiver for multi-modal data is still a benchmark to achieve. To this end, we propose a multi-modal neural receiver based end to end system that not only efficiently estimates the channel, but also improves the performance metrics such as peak signal-to-noise ratio, mean square error etc to reconstruct multi sensors’ data. 
This allows for accurate decision-making in real-time scenarios, such as vehicle navigation, obstacle detection, and human machine interactions. 

\begin{figure*}[t]
    \centering
    \includegraphics[width=1\textwidth,height=0.25\textheight]{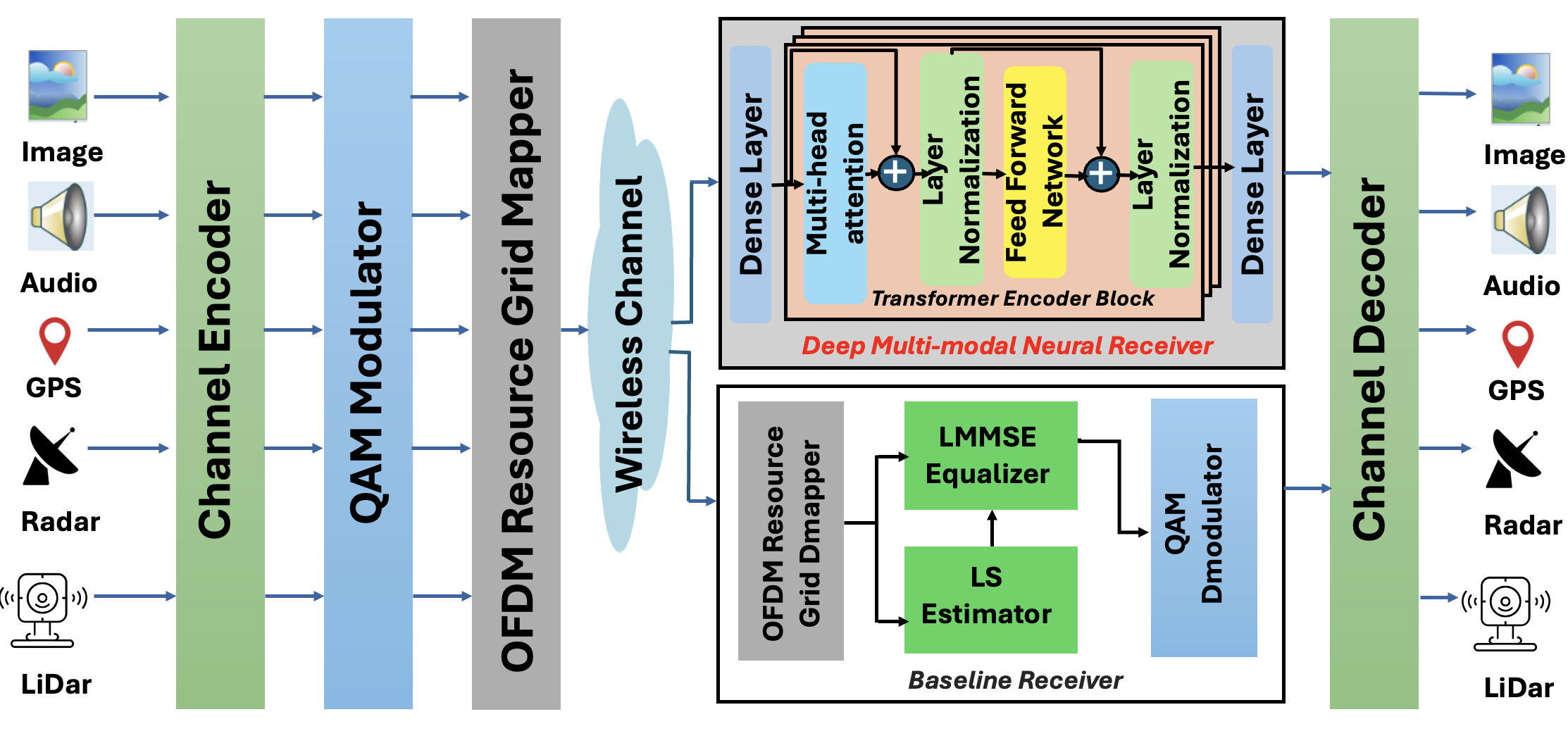}
    \caption{\textbf System Model}
    \label{fig1}
\end{figure*}

Our main contributions include:
\begin{enumerate}
    \item We identify key training parameters in a neural receiver that influences the reliability.
    \item We evaluate multiple neural receiver architectures and propose best fit model for large scale deployment.
    \item We conduct multi-modal analysis of the proposed neural receiver for V2X data using simulation scenarios.
\end{enumerate}

The rest of the paper is structured as follows: Section~\ref{s3} provides a detailed description of the physical layer design for both the transmitter and receiver. This section covers the functionalities of a system with a conventional receiver as well as with neural receiver. Section~\ref{s4} outlines the experimental setup that has been followed for testing, with the multi-modal results discussed in Section~\ref{s5}. Lastly, Section~\ref{s6} provides the conclusion and suggest future directions.
\section{System Model} \label{s3}

This section present the PHY layer functionalities of transmitter and receiver. We assume that an Automated Vehicle (AV) communicates with a central base station (BS) by transmitting its data. As shown in Fig.~\ref{fig1}, when information bits reach the vehicle's physical layer (transmitter), they are initially encoded using channel encoder. After encoding, the data is modulated to produce baseband symbols. These symbols are then passed through an Orthogonal Frequency Division Multiplexing (OFDM) resource grid mapper, where known pilot symbols are added to aid the receiver in estimating the wireless channel conditions. 
Once the cyclic prefix is added, the signal is transmitted by an AV and it goes through a wireless channel, where it undergoes various distortions such as noise, fading, and interference. 

In the following subsections, we present the baseline PHY layer of the receiver (i.e., BS) in subsection \ref{s31} and the neural receiver-based PHY layer is presented in subsection \ref{s32}. 

\subsection{Baseline Receiver} \label{s31}
At BS, the first step is to estimate the channel using Least Square (LS) estimator as:
\begin{equation}
\mathbf{\hat{H}} = \arg \min_{\mathbf{H}} \|\mathbf{Y} - \mathbf{H} \mathbf{X}\|_2^2    
\end{equation}

Where:
\begin{itemize}
    \item $\mathbf{Y}$ is the received signal vector,
    \item $\mathbf{X}$ is the transmitted symbol vector,
    \item $\mathbf{H}$ is the channel estimate matrix,
    \item $\|\cdot\|_2^2$ is the squared $L_2$-norm
\end{itemize}

As a next step, the received signal is equalized using:
\begin{equation}
\mathbf{\hat{X}} = \mathbf{W} \mathbf{Y}    
\end{equation}

Where $\mathbf{W}$ is the equalizer matrix, optimized using the minimum mean squared error (MMSE) criterion:
\begin{equation}
\mathbf{W}_{\text{MMSE}} = \left(\mathbf{H}^H \mathbf{H} + \sigma_n^2 \mathbf{I}\right)^{-1} \mathbf{H}^H    
\end{equation}

Where:
\begin{itemize}
    \item $\mathbf{H}^H$ is the Hermitian transpose of the channel matrix,
    \item $\sigma_n^2$ is the noise variance,
    \item $\mathbf{I}$ is the identity matrix.
\end{itemize}

After equalization, it is demodulated to compute the log-likelihood ratio (LLR) using:
\begin{equation}
LLR(b_i) = \log \left( \frac{P(b_i = 1 | \mathbf{\hat{X}}, \mathbf{H})}{P(b_i = 0 | \mathbf{\hat{X}}, \mathbf{H})} \right)
\end{equation}

Where:
\begin{itemize}
    \item $P(b_i = 1 | \mathbf{\hat{X}}, \mathbf{H})$ and $P(b_i = 0 | \mathbf{\hat{X}}, \mathbf{H})$ are the probabilities of bit $b_i$ being 1 or 0, given the equalized symbols, channel estimates, and receiver parameters.
\end{itemize}

At the last step, this demodulated signal goes through a channel decoder to reconstruct original message sent by an AV using:
\begin{equation}
\mathbf{\hat{b}} = \arg \min_{\mathbf{b}} \sum_{i=1}^{N_b} \left( b_i \cdot LLR(b_i) \right)    
\end{equation}

Where:
\begin{itemize}
    \item $N_b$ is the total number of transmitted bits.
    \item $\mathbf{b} = [b_1, b_2, \dots, b_{N_b}]$ is the bit sequence.
\end{itemize}

\subsection{Neural Receiver} \label{s32}


Our proposed neural receiver processes the baseline receiver functionalities to recover transmitted data. It gets received signal as input, and computes optimal LLRs at the output. The architecture consists of an input dense layer which is followed by multiple transformer encoder blocks. Transformer encoder blocks operate on multihead self attention mechanism as its core learning algorithm \cite{11}. Mathematically, it is presented as:

\begin{equation}
\text{Attention}(Q, K, V) = \text{softmax}\left(\frac{QK^T}{\sqrt{d_k}}\right)V
\end{equation}

where $Q$, $K$, and $V$ represent the query, key, and values, respectively. Multi-head attention operates by executing several attention processes simultaneously, each with its unique learned projections. The result of these parallel processes are merged and subjected to a linear transformation to generate the ultimate output. This method enables the model to simultaneously process information from various representational subspaces across different positions. Once the signal is processed by transformer encoder blocks, it goes through an output dense layer to generate optimal LLRs.

The proposed neural receiver model is trained using a supervised learning approach, where the objective is to minimize the error between transmitted and reconstructed information bits;

\begin{equation}
\text{BER} = \frac{1}{N_b} \sum_{i=1}^{N_b} P(\hat{b}_i \neq b_i)    
\end{equation}

Where:
\begin{itemize}
    \item $N_b$ is the total number of transmitted bits,
    \item $b_i$ is the transmitted bit,
    \item $\hat{b}_i$ is the estimated LLR,
    \item $P(\hat{b}_i \neq b_i)$ is the probability of error for bit $i$.
\end{itemize}

It takes noisy OFDM symbols as input along with the noise power, which helps in improving robustness. The architecture utilizes multiple transformer blocks to extract complex patterns from the received signal and produce log-likelihood ratios (LLRs) for channel decoding. The output of the neural receiver is compared against the actual transmitted bits, and the training is done using the binary cross-entropy (BCE) loss:

\begin{table}[b]
    \centering
    \caption{Proposed Neural Receiver Parameters}
    \begin{tabular}{|c|c|} \hline
        \textbf{Parameter} & \textbf{Value} \\ \hline
        Feed Forward Network Dimension & 128 \\ \hline
        Embedding Dimension & 128 \\ \hline
        Learning Rate & 1$e^{-3}$  \\ \hline
        Optimizer & AdamW \\ \hline
        Activation Function & Relu \\ \hline
        Training Batch Size & 32 \\ \hline
        Number of Training Iteration & 375,000 \\ \hline
    \end{tabular}  
    \label{table2}
\end{table}

\begin{equation}
\mathcal{L}_{\text{BCE}} = - \frac{1}{N} \sum_{i=1}^{N} \left[ b_i \log(\sigma(\hat{b}_i)) + (1 - b_i) \log(1 - \sigma(\hat{b}_i)) \right]    
\end{equation}

Where:
\begin{itemize}
    \item $N$ is the total number of bits,
    \item \( b_i \) is the actual bit,
    \item $\hat{b}_i$ is the predicted bit,
    \item \( \sigma \) is the sigmoid activation function
\end{itemize}



The proposed neural receiver model is trained on sionna \cite{8} with $37500$ iterations and a batch size of $32$ as shown in Table.~\ref{table2}. Each iteration involves randomly sampling an SNR value which is followed by generating random channel topologies using a 3GPP Urban Macro (UMa) channel model as specified in \cite{9}. This introduces realistic and diverse channel conditions to improve the generalizability of the model. 
Metrics like accuracy and Binary Cross Entropy (BCE) is monitored to track the model's convergence and effectiveness in predicting the transmitted bits accurately.

\section{Implementation and Experimental Setup} \label{s4}


\begin{table}[t]
    \centering
    \caption{Wireless Communication Parameters}
    \begin{tabular}{|c|c|} \hline
        \textbf{Parameter} & \textbf{Value} \\ \hline 
        Carrier Frequency & 28GHz \\ \hline
        Physical Channel & UMa \\ \hline
        Modulation & 64 QAM \\ \hline
        Code rate & 0.5 \\ \hline
        Subcarrier Spacing & 240KHz \\ \hline
        No. of Transmitter Antenna & 1 \\ \hline
        No. of Receiver Antenna & 2 \\ \hline
        No. of OFDM symbol & 14 \\ \hline
        Fast Fourier Transform Size & 128 \\ \hline
        Minimum Vehicle Speed & 60 km/h \\ \hline
        Maximum Vehicle Speed & 120 km/h \\ \hline
    \end{tabular}
    \label{table1}
\end{table}

To evaluate our proposed neural receiver based end to end setup, we assume a 
communication system that uses an OFDM-based waveform with $128$ subcarriers and a subcarrier spacing of $240$ kHz as shown in Table.~\ref{table1}. The resource grid is structured with $14$ OFDM symbols per frame, incorporating a Kronecker pilot pattern for aiding channel estimation. Each grid carries $64$-QAM-modulated data, encoding $6$ bits per symbol. This grid is transmitted by the vehicle and it goes through a UMa channel model. The channel simulates real-world wireless environments, including path loss and shadow fading in uplink environment. At BS, we first build our proposed neural receiver model and apply weights to it. It then processes the received resource grid for inference to compute LLR. These LLRs are then processed by LDPC decoder to evaluate the performance metrics between the transmitted data and reconstructed data. The steps to evaluate our propose neural receiver model is shown in Algorithm.~\ref{algo1}, where we first identify the key parameters that influences reliability and then conduct multi-modal analysis of the best fit model discussed in Section.~\ref{s5}. For evaluation of multi-modal V2X data, we consider 3 performance metrics, namely:

\begin{itemize}
    \item Peak Signal to Noise Ratio (PSNR)
    \item Mean Squared Error (MSE)
    \item Root Mean Squared Error (RMSE)
\end{itemize}



\section{Results and Discussion} \label{s5}

\begin{algorithm}[t] 
\caption{Proposed Neural Receiver Model Evaluation}
\label{algo1}
\begin{algorithmic}[1]
\State \textbf{\textit{\#Input}}: 
\newline Trained neural receiver models
\newline Generic testing data
\newline Multi-modal testing data
\State Filter the number of transformer blocks by balancing performance and architecture complexity
\State Select the number of attention heads
\State Define evaluation metrics for different data types
\State Multi-modal assessment of neural receiver performances
\State \textbf{\textit{\#Output: }}
\newline Characterization of the best fit model
\end{algorithmic}
\end{algorithm}

\begin{figure}[t]
  \centering
  \includegraphics[width=0.45\textwidth,height=0.25\textheight]{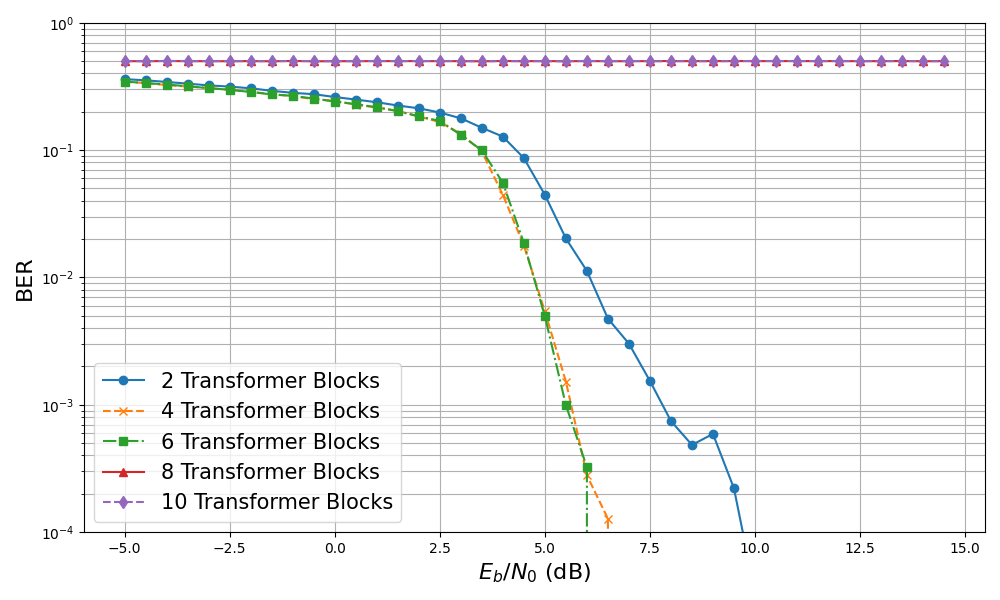}
  \caption{Comparison of BER w.r.t SNR by varying transformer encoder blocks of neural receiver}
  \label{fig3}
\end{figure}

This section is divided into two parts. In the first part, we present end to end performance by changing the trainable parameters of the model to analyze and retrieve the best fit parameters. In the next step, we present the results of neural receiver architecture with best fit trainable parameters on multi-modal data. 

Fig.~\ref{fig3} shows the comparison of BER with respect to (w.r.t) SNR for multiple neural receivers by varying the number of transformer encoder blocks in each neural receiver. We assume an AV operating at an average speed of 90 km/hr, with the signal received at BS propagating through an UMa channel. As shown in the figure, the neural receiver model with 2 transformer encoder blocks converges to minimal BER at comparatively higher SNR. Similarly, the model diverges with $8$ and $10$ blocks. However, it shows good results with $4$ and $6$ blocks. Since both $4$ and $6$ blocks architecture shows close result to each other, we select the neural receiver model with 4 transformer encoder blocks for next simulations to reduce architecture complexity.

\begin{figure}[t]
  \centering
  \includegraphics[width=0.45\textwidth,height=0.25\textheight]{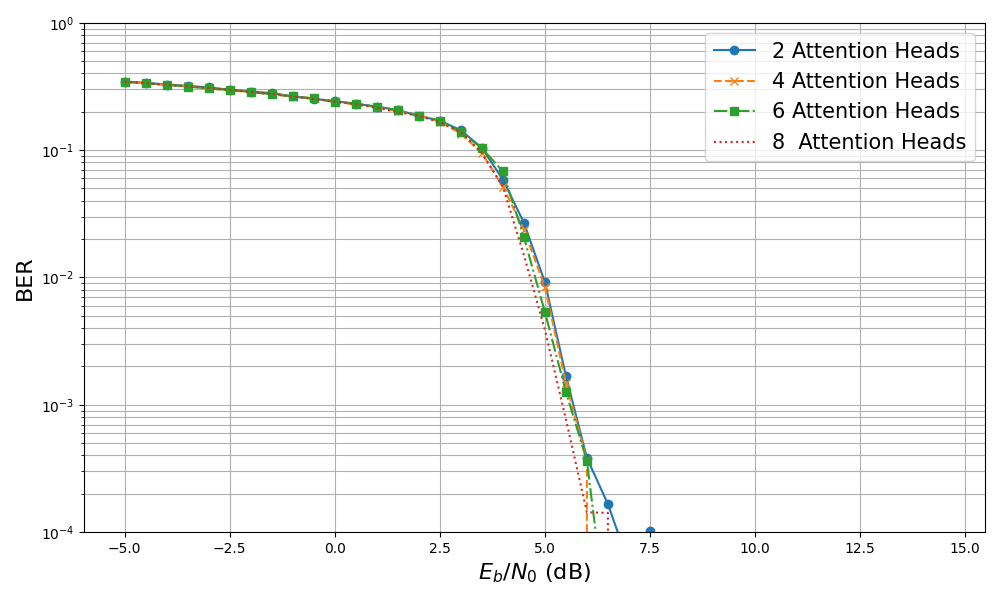}
  \caption{Comparison of BER w.r.t SNR by varying attention heads of neural receiver}
  \label{fig4}
\end{figure}

Later, we evaluate the performance of our proposed neural receiver by varying the number of attention heads inside the transformer encoder blocks. The result is shown in Fig.~\ref{fig4} where we observe minimal variation in BER; however, the model with encoder blocks having 8 attention heads demonstrates best performance. Therefore, we propose a neural receiver with 4 transformer encoder blocks and 8 attention heads in each block for large-scale deployment.

\begin{figure*}[t]
    \centering
    \begin{subfigure}{0.20\textwidth}
        \centering
        \includegraphics[width=\textwidth]{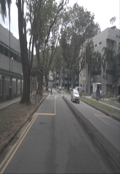}
        \caption{Vehicle \\ Transmitted Image}
        \label{fig71}
    \end{subfigure}%
    \begin{subfigure}{0.20\textwidth}
        \centering
        \includegraphics[width=\textwidth]{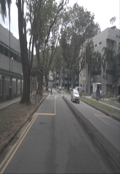}
        \caption{Proposed \\ Reconstructed Image}
        \label{fig72}
    \end{subfigure}%
    \begin{subfigure}{0.20\textwidth}
        \centering
        \includegraphics[width=\textwidth]{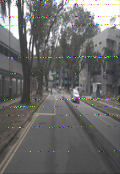}
        \caption{DeepRx \\ Reconstructed Image}
        \label{fig73}
    \end{subfigure}%
    \begin{subfigure}{0.20\textwidth}
        \centering
        \includegraphics[width=\textwidth]{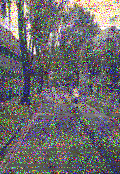}
        \caption{Baseline \\ Reconstructed Image}
        \label{fig74}
    \end{subfigure}
    \caption{Image taken by an AV and transmitted to the BS with speed of $90$km/hr over UMa channel}
    \label{fig55}
\end{figure*}

\subsection{Comparison}

To evaluate the performance of our proposed neural receiver on multi-modal V2X data obtained from DeepSense 6G \cite{12}, we compare it with $2$ state of the art approaches:

\begin{itemize}
    \item \textbf{Baseline}, this evaluates the channel conditions using known data symbols and subcarriers added during transmission. Later, it demodulates the signal using QAM demodulator.
    \item \textbf{DeepRx}, this architecture operates on convolution as its core learning algorithm. We re-trained the convolutional neural network (CNN) model from \cite{6} using the same data and channel model which allowed us to directly evaluate the performance differences between the two architectures under identical conditions.
\end{itemize}

\subsubsection{Image Transmission use case}

\begin{figure}[t]
  \centering
  \includegraphics[width=0.45\textwidth,height=0.25\textheight]{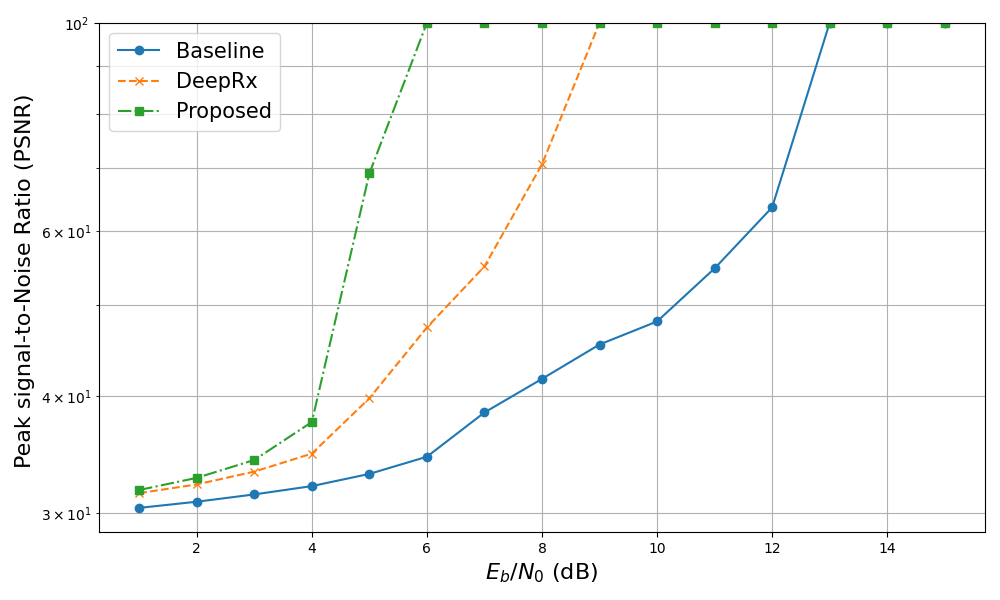}
  \caption{Comparison of PSNR w.r.t SNR for image transmitted over UMa Channel Model with vehicle speed ranging from 60 to 120 km/h}
  \label{fig5}
\end{figure}

\begin{figure}[t]
  \centering
  \includegraphics[width=0.45\textwidth,height=0.25\textheight]{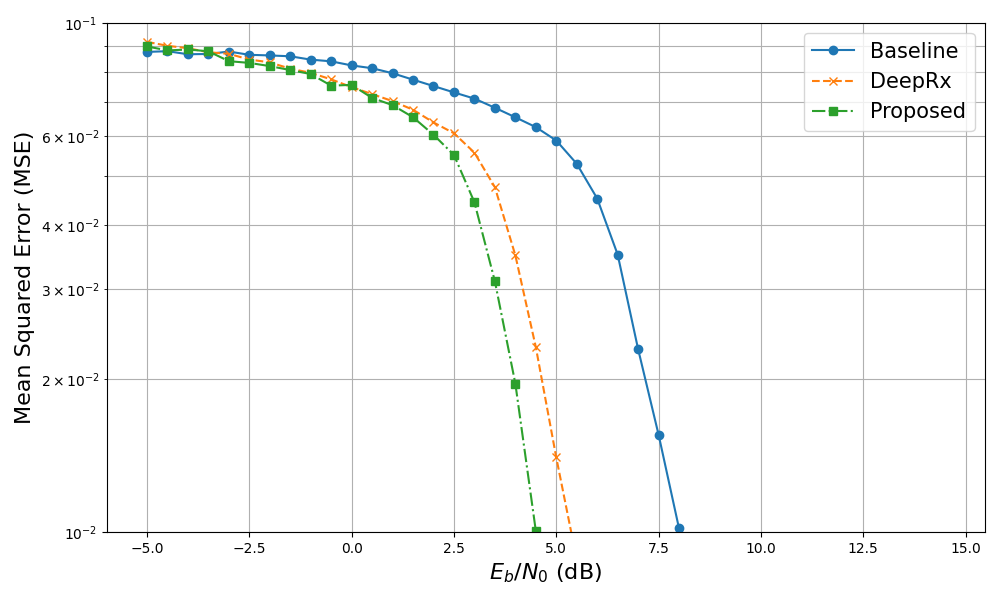}
  \caption{Comparison of MSE w.r.t SNR for audio transmitted over UMa Channel Model with vehicle speed ranging from 60 to 120 km/h}
  \label{fig6}
\end{figure}

We evaluate the performance of our proposed neural receiver model compared to the state of the art approaches for image transmission use case. We assume that an AV is moving with a speed of $90$km/hr, and it captures an image. This image is transmitted by AV and it goes through UMa channel with SNR of $6$dB. At BS, it is reconstructed by our proposed neural receiver, DeepRx and baseline, and the results are shown in Fig.~\ref{fig55}. Our proposed neural receiver achieves superior image reconstruction quality with fewer missing pixels which is followed by DeepRx. In contrast, the Baseline exhibits worst performance in reconstructing the image.

To quantify the results, we compute Peak Signal-to-noise Ratio (PSNR) w.r.t SNR. PSNR measures the quality of a reconstructed image by comparing the maximum possible signal strength to the noise introduced by the reconstruction process. Results are shown in Fig.~\ref{fig5} which demonstrates that as SNR increases, PSNR also increases because higher PSNR value indicates better reconstruction quality. It can be seen that our proposed neural receiver reconstruct image similar to the one transmitted by an AV at $6$dB SNR. However, Baseline and DeepRx achieve the same at $9$dB and $13$dB SNR respectively.

\begin{figure}[t]
  \centering
  \includegraphics[width=0.45\textwidth,height=0.25\textheight]{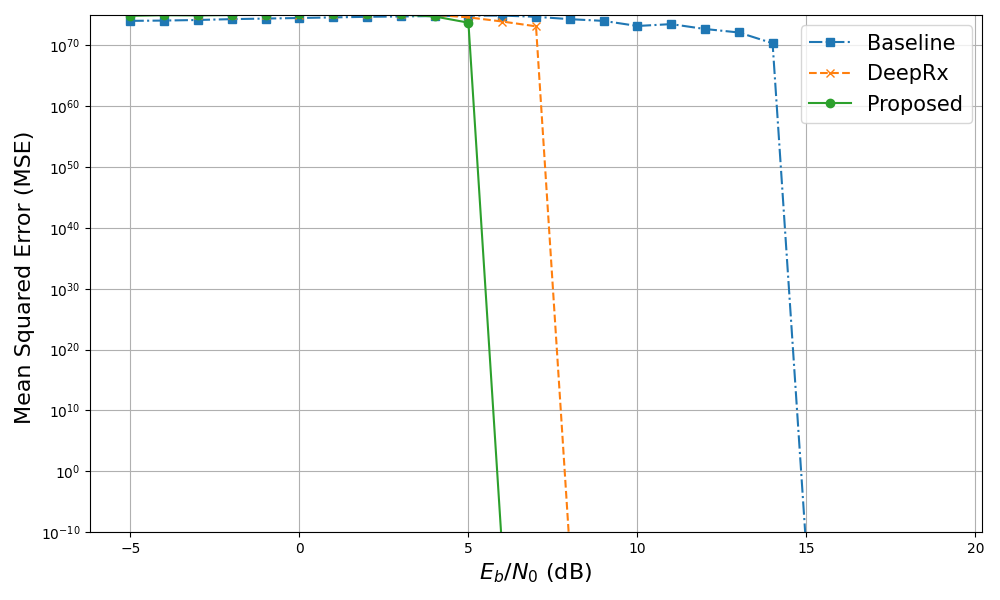}
  \caption{Comparison of MSE w.r.t SNR for LiDar cloud points transmitted over UMa Channel Model with vehicle speed ranging from 60 to 120 km/h}
  \label{fig7}
\end{figure}

\subsubsection{Audio Transmission use case}
In order to evaluate our proposed neural receiver for use cases involving audio transmission, we computed Mean Squared Error (MSE) between the transmitted and reconstructed audio for SNR ranging from $-5$ dB to $15$ dB. The results are shown in Fig.~\ref{fig6} where it can be seen that our proposed neural receiver reaches near-zero MSE at $4.5$ dB, demonstrating an improvement of $1$ dB in SNR over DeepRx and $3.5$ dB over the Baseline.

\subsubsection{GPS Transmission use case}
We also evaluate our proposed neural receiver for gps coordinates transmission use case. We compute Root Mean Squared Error (RMSE) between transmitted and reconstructed coordinates and results are shown in Table~\ref{table3}. It can be seen that our proposed neural receiver successfully reconstructs the transmitted coordinates at minimal SNR of $5.4$ dB, while DeepRx and Baseline achieve the same at $6.4$ dB and $7$ dB respectively. We observe that our proposed neural receiver achieves zero RMSE at SNR improvement of $1$ dB and $1.6$ dB over DeepRx and Baseline respectively.

\begin{table}[t]
    \centering
    \caption{Minimum SNR to reach zero RMSE for GPS Reconstruction}
    \begin{tabular}{|c|c|} \hline
        \textbf{Algorithm} & \textbf{SNR (dB)} \\ \hline
        Proposed & 5.4 \\ \hline
        DeepRx & 6.4 \\ \hline
        Baseline & 7 \\ \hline
    \end{tabular}  
    \label{table3}
\end{table}

\subsubsection{LiDAR Cloud Points Transmission use case} 
In order to evaluate our proposed neural receiver for LiDAR cloud points transmission use case, we compute MSE between the transmitted and reconstructed cloud points by varying the SNR values. The results are shown in Fig.~\ref{fig7} and it can be seen that our proposed neural receiver outperforms both the state-of-the-art Baseline and DeepRx. It achieves zero MSE at SNR of $6$ dB, while DeepRx and Baseline achieve the same at $8$ dB and $15$ dB, respectively. 

\subsubsection{Radar Detection Transmission use case}
\begin{figure}[t]
  \centering
  \includegraphics[width=0.45\textwidth,height=0.25\textheight]{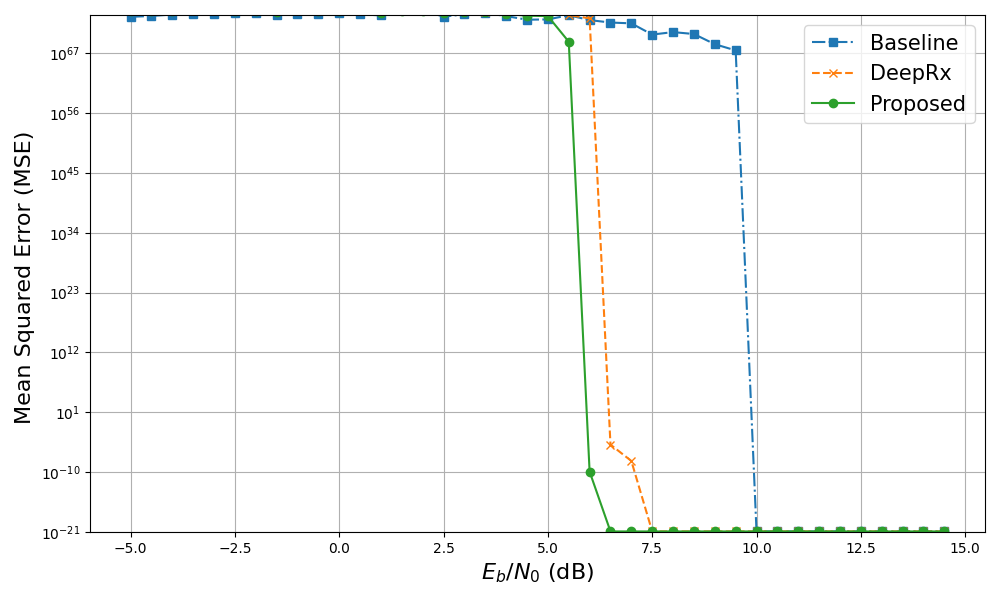}
  \caption{Comparison of MSE w.r.t SNR for radar signal transmitted over UMa Channel Model with vehicle speed ranging from 60 to 120 km/h}
  \label{fig8}
\end{figure}

We also evaluate the performance of our proposed neural receiver for radar detection use case by calculating MSE between transmitted and reconstructed signal at various SNR levels. The results are shown in Fig.~\ref{fig8} which shows that our proposed neural receiver model consistently outperform both the baseline and DeepRx. Specifically, it achieves near-zero MSE at an SNR of $6$ dB, whereas DeepRx and baseline reaches similar performance at $7.5$ dB and $10$ dB, respectively. 

\section{Conclusion and Future Work} \label{s6}
In this paper, we proposed a multi-modal neural receiver designed to provide hyper reliability for large-scale deployment. Our proposed neural receiver takes the received resource grid as input and generates optimal Log Likelihood Ratio (LLR) at the output, replacing traditional component that includes LS estimator, LMMSE equalizer, and QAM demodulator. We conduct a comprehensive multi-modal analysis, testing our model's performance on various V2X data flows, including image, audio, GPS, LiDAR, and radar. Simulation results show that our proposed neural receiver consistently outperforms the state-of-the-art Baseline and DeepRx across all tested V2X domain data.

In future work, we plan to conduct a detailed latency analysis of existing neural receivers to assess their efficiency in large scale data and real-time experiments.

\end{document}